\documentclass[11pt,a4paper]{article}
\usepackage[top=30truemm,bottom=30truemm,left=25truemm,right=25truemm]{geometry}
\usepackage[]{graphicx}
\usepackage{here}
\usepackage{bm}
\usepackage{amsmath}
\begin{document}

\begin{center} {\LARGE Collective ratchet transport generated by particle crowding under asymmetric sawtooth-shaped static potential}
\end{center}
\centerline {Masayuki Hayakawa$^{a,b}$, Yusuke Kishino$^b$, and Masahiro Takinoue$^{b,c,\ast}$}
\vspace {0.1in}
\noindent {$^a$RIKEN Center for Biosystems Dynamics Research, Kobe, Hyogo 650-0047, Japan; $^b$Department of Computational Intelligence and Systems Science, Tokyo Institute of Technology, Yokohama, Kanagawa 226-8502, Japan; $^c$Department of Computer Science, School of Computing, Tokyo Institute of Technology, Yokohama, Kanagawa 226-8502, Japan}

\noindent{$^\ast$E-mail:takinoue@c.titech.ac.jp}

\vspace {0.5in}

\leftline {\large \textbf{Abstruct}}

In this study, we describe the ratchet transport of particles under static asymmetric potential with periodicity. Ratchet transport has garnered considerable attention due to its potential for developing smart transport techniques on a micrometer scale. In previous studies, either particle self-propulsion or time varying potential was introduced to realize unidirectional transport.
Without utilizing these two factors, we experimentally demonstrate ratchet transport through particle interactions during crowding.
Such ratchet transport induced by particle interaction has not been experimentally demonstrated thus far, although some theoretical studies had suggested that particle crowding enhances ratchet transport.
In addition, we constructed a model for such transport in which the potential varies depending on the particle density, which agrees well with our experimental results. This study can accelerate the development of transport techniques on a micrometer scale.

\vspace {0.5in}

\leftline {\large \textbf{Introduction}}
Increasing attention has been focused on techniques for transporting micrometer-sized particles using asymmetric periodic potential \cite{reimann2002brownian,astumian1997thermodynamics,lau2017introduction,
caballero2016motion,reichhardt2017ratchet}.
Generally, such transport is called ratchet transport, where an asymmetric ratchet-shaped potential rectifies particle motion, resulting in unidirectional particle motion, even if the particles intrinsically exhibit nondirectional motion.
Thus, ratchet transport enables unidirectional transport without macroscopic field gradients, including microflow and electric potential, along the transport direction, and is expected to be applied for smart transportation methods in lab-on-a-chip devices \cite{caballero2016motion,reichhardt2017ratchet}.
To date, ratchet transport for various types of particles have been proposed including the transport of self-propelled particles (ex. migrating cells) and passive (nonself-propelled) particles (ex. molecules) \cite{reimann2002brownian,astumian1997thermodynamics,lau2017introduction,
caballero2016motion,reichhardt2017ratchet}.
Of late, techniques to rectify the motion of self-propelled particles are being developed using static asymmetric potential.
Generally, although self-propelled particles such as migrating cells and micromotors exhibit ballistic motion on a short timescale, they lose directionality after the rotational diffusion time \cite{howse2007self}.
Thus, the motion of such self-propelled particles must be guided.
Solid barriers fabricated using lithography were frequently used in previous studies as the static asymmetric potential \cite{koumakis2013targeted,katuri2018directed,hiratsuka2001controlling}.
Microparticles suspended in a bath of swimming bacteria were driven by collisions with the bacteria, and their nondirected motions were rectified using three-dimensionally fabricated patterned asymmetric solid barriers \cite{koumakis2013targeted}.
Furthermore, the unidirectional motions of catalytic micromotors and kinesin-driven microtubules were successfully extracted using arrowhead-shaped microchannels \cite{katuri2018directed,hiratsuka2001controlling}.

On the other hand, techniques to rectify the motion of passive particles have been studied, using dynamic asymmetric potential \cite{rousselet1994directional,faucheux1995optical,bader1999dna,skaug2018nanofluidic}.
Although the particles themselves do not exhibit ballistic motion, unidirectional motion is extracted through the time-dependency and asymmetricity of the potential.
Typical examples for varying the potential have been proposed \cite{reimann2002brownian,astumian1997thermodynamics,lau2017introduction}; in a flashing ratchet, the asymmetric potential switches between the on/off states, whereas in a rocking ratchet, the asymmetric potential is tilted by an oscillating force.
Based on these principals, it has been experimentally demonstrated that passive small particles can be transported.
For instance, thermally fluctuating colloidal particles \cite{rousselet1994directional,faucheux1995optical} and large DNA molecules \cite{bader1999dna} were transported by the on/off switching of asymmetric potential.
In addition, a recent study has realized the transportation of gold spheres under asymmetric potential tilted by an applied oscillating electric field \cite{skaug2018nanofluidic}.

Regarding ratchet transport, several theoretical studies have reported that interaction during particle crowding causes nontrivial and complex effects, for example, the reinforcement of the transport flux.
This phenomenon generated by crowding has been theoretically observed in the transport of passive particles under dynamic potential as well as self-propelled particles under static potential.
As an example, one-dimensional models which forbid the overlapping of particles show that particle interaction enhances the transport of passive particles under dynamic potential \cite{derenyi1995cooperative,derenyi1996collective,csahok1997transport,zheng2010cooperative}.
Furthermore, examination through computer simulation of the behavior of sterically interacting self-propelled run-and-tumble particles in the presence of a quasi-one-dimensional asymmetric substrate indicates that the particle flux can be increased \cite{mcdermott2016collective}.

In this study, we experimentally demonstrate the ratchet transport of passive particles under static asymmetric potential, utilizing the effect of particle interaction. Here, interacting passive microparticles are transported under a static asymmetric electric potential generated using two-dimensional sawtooth-shaped electrodes. In this system, the transport occurs depending on the microparticle density, as predicted by theoretical studies \cite{derenyi1995cooperative,derenyi1996collective}. We believe that our study can enhance the development of smart applications in lab-on-a-chip devices and electrically controlled molecular robots.

\vspace {0.5in}

\begin{figure}[t]
\centering
\includegraphics[scale=0.9]{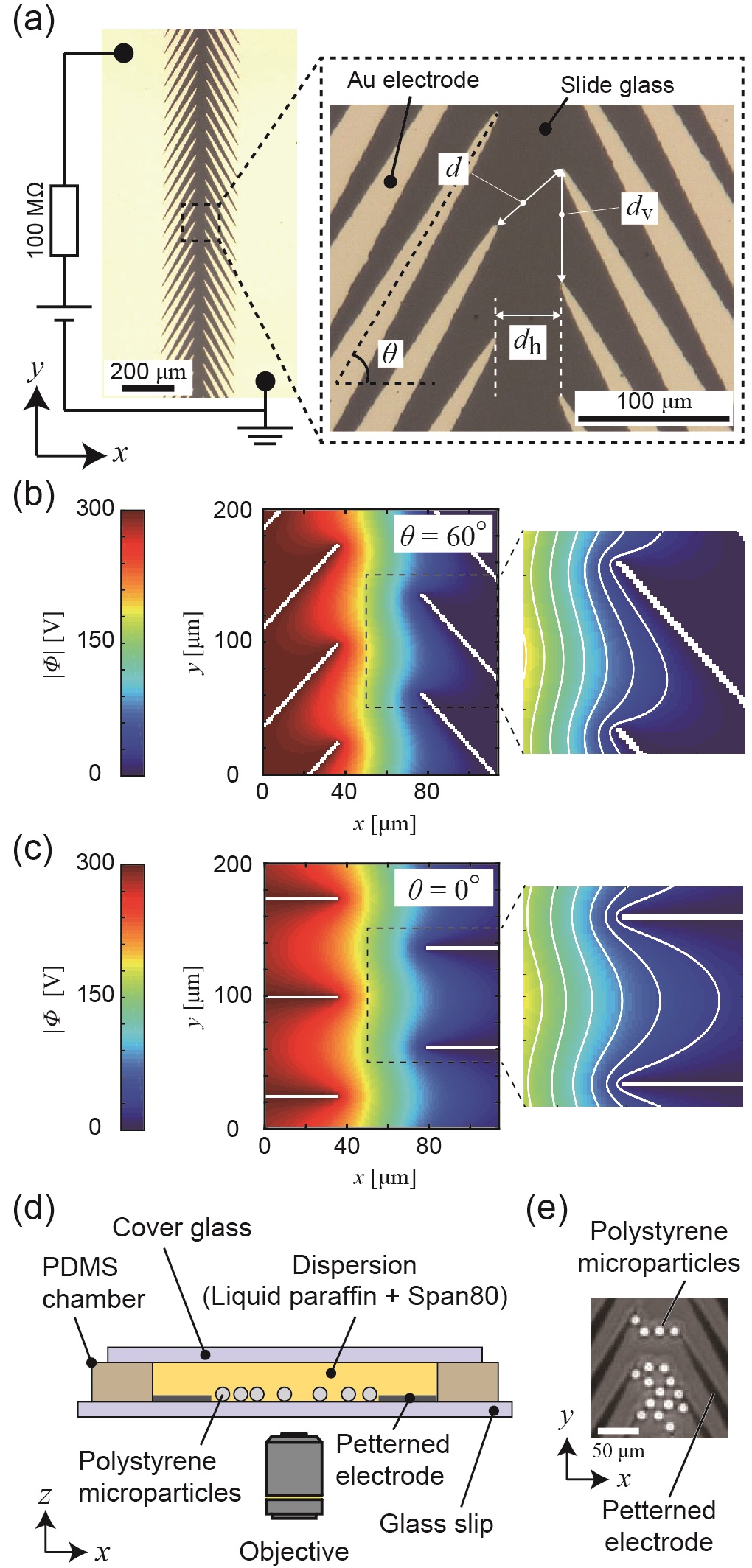}
\caption{
Experimental setup and calculated electric potential. (a) Sawtooth-shaped Au microelectrode. The electrodes were two-dimensionally patterned on a glass slide. In all the experiments, $d$, $d_\mathrm{h}$, and $d_\mathrm{v}$ were set to 53 $\mu$m, 40 $\mu$m, and 70 $\mu$m, respectively. The angle of the sawtooth $\theta = 60^{\circ}$. (b, c) Calculated electric potential $|\it\Phi|$. The white lines in the magnified images are the contours of the potentials. (b) Asymmetric potential generated by the electrodes at $\theta = 60^{\circ}$. (c) Symmetric potential generated by the electrodes at $\theta = 0^{\circ}$. (d) Experimental setup for the observation of microparticle motion. Microparticles dispersed in liquid paraffin were observed in the presence of the electric potential generated by the sawtooth-shaped electrodes. (e) Typical top-view image of experiments.
}
\label{fig:SU}
\end{figure}

\leftline {\large \textbf{Results and Discussion}}
\vspace {0.1in}
\noindent \textbf{Microparticle motion in a sparsely populated area.} We generated an electric field using sawtooth-shaped Au microelectrodes patterned on a two-dimensional $xy$-plane (Fig.  \ref{fig:SU}a). A constant direct current (DC) power was supplied to the microelectrode (the applied voltage $\phi$ was $\sim$ 300 V); i.e., the electric field was stable over time $t$. In all the experiments, the electrode on the left was positive, whereas the one on the right was negative.
The direct distance $d$, horizontal distance $d_\mathrm{h}$, and vertical distance $d_\mathrm{v}$ between the tips of the sawtooth were 53 $\mathrm{\mu}$m, 40 $\mathrm{\mu}$m, and 70 $\mathrm{\mu}$m, respectively (Fig. \ref{fig:SU}a).
The asymmetricity of the field along the $y$-axis was varied by changing the angle of the sawtooth $\theta$; $\theta = 60^{\circ}$ (Fig. \ref{fig:SU}a and Fig. S1a), $\theta = 30^{\circ}$ (Fig. S1b), and $\theta = 0^{\circ}$ (Fig. S1c).
Figs. \ref{fig:SU}b and \ref{fig:SU}c show the calculated electric potential $|\it\Phi|$ when $\theta = 60^{\circ}$ and $\theta = 0^{\circ}$.
In both cases, total gradint alogn the $y$-axis is zero.
The symmetricity of $|\it\Phi|$ at $\theta = 60^{\circ}$ and $30^{\circ}$ is broken along the $y$-axis (Fig. \ref{fig:SU}b and Fig. S2), whereas $|\it\Phi|$ at $\theta = 0^{\circ}$ is symmetric along the $y$-axis (Fig. \ref{fig:SU}c).
The asymmetricity of $|\it\Phi|$ is remarkable near the electrode.
Similarly, the calculated electric fields are asymmetric when $\theta = 60^{\circ}$ and $\theta = 30^{\circ}$ (Fig. S3).

In this study, we observed the motion of polystyrene microparticles (20 $\mathrm{\mu}$m diameter) between the microelectrodes.
These microparticles were dispersed in liquid paraffin containing 0.1$\% \mathrm{(w/w)}$ Span80 (sorbitan monooleate), filled in a polydimethylsiloxane (PDMS) chamber ($\sim$1 mm height) placed on a glass slide with sawtooth-shaped Au electrodes (Fig. \ref{fig:SU}d).
Fig. \ref{fig:SU}e depicts the typical top-view image of experiments.

\begin{figure*}[!t]
\centering
\includegraphics[scale=0.25]{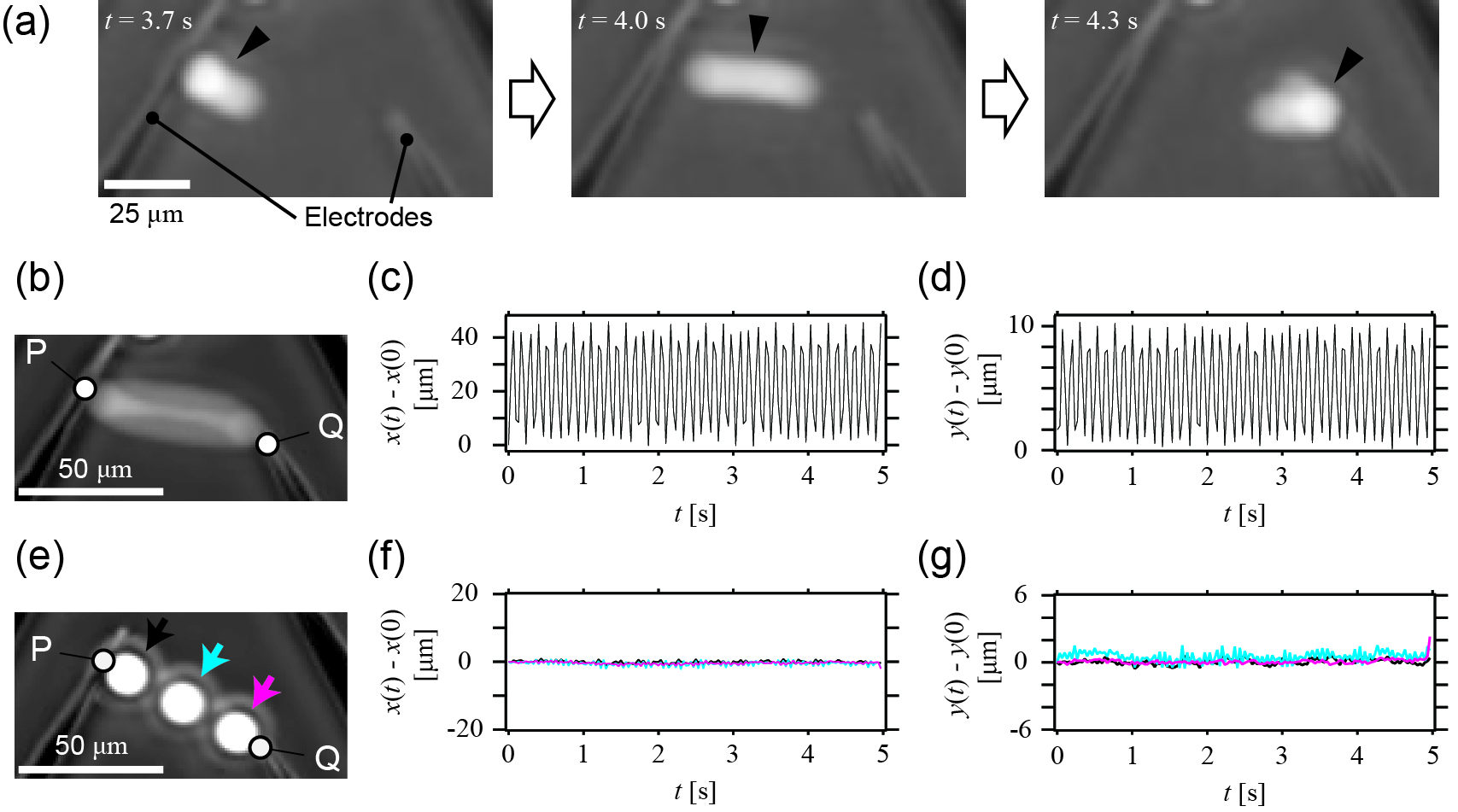}
\caption{
Motion of microparticles confined in between the tip of the sawtooth. (a) Time-lapse images of the motion of a single microparticle. The microparticle is indicated by arrowheads. (b) Maximum intensity projection image of the motion of the single microparticle depicted in (a). Back-and-forth motion was generated on the line between points P and Q. (c, d) Time variations of the $x$ and $y$ displacements of the back-and-forth motion. (e) Maximum intensity projection image of the three microparticles aligned between the tip of the sawtooth. The microparticles were statically aligned on line PQ. (f, g) Time variations of the $x$ and $y$ displacements of the three microparticles. The color of each line corresponds to the color of the arrow in (e).
}
\label{fig:SPM}
\end{figure*}

Fig. \ref{fig:SPM}a shows the motion of a single microparticle (indicated by arrowheads) between the tips of the electrodes (Movie S1).
As shown in the maximum intensity projection (Fig. \ref{fig:SPM}b) and the relative position from the initial position, $x(t)-x(0)$ and $y(t)-y(0)$ (Figs. \ref{fig:SPM}c and \ref{fig:SPM}d), the microparticles exhibited back-and-forth oscillatory motion between the two points P and Q depicted in Fig. \ref{fig:SPM}b.
We speculate that this back-and-forth motion can be attributed to a combination of electrophoretic and dielectrophoretic forces \cite{takinoue2010rotary,kurimura2013back,bishop2018contact}, which is the same mechanism we had previously reported \cite{takinoue2010rotary}.
When DC voltage was applied to the electrodes, the microparticles were attracted to the tip of the sawtooth (point Q) by the dielectrophoretic force, obtaining electric charge.
Immediately after obtaining the charge, the microparticles were repelled from the tip by electrostatic repulsion and attracted to the opposite electrode (point P). After the electrostatic repulsion from point P, the microparticles were attracted to the opposite tip (point Q) again by the dielectrophoretic force.
In the case, where there are two microparticles between the electrodes, the back-and-forth motions were observed as well as the single microparticle (Fig. S4).

Fig. \ref{fig:SPM}e shows the maximum intensity projection of Movie S2, which depicts the case, that there were three microparticles between the electrodes.
The time variations $x(t)-x(0)$ and $y(t)-y(0)$ of each microparticle are shown in Figs. \ref{fig:SPM}f and \ref{fig:SPM}g.
The color of the arrows in Fig. \ref{fig:SPM}e, and lines in Figs. \ref{fig:SPM}f and \ref{fig:SPM}g identify each microparticle.
In this situation, the three microparticles were aligned between points P and Q due to dielectric polarization; therefore, the $x$- and $y$-coordinates of each microparticle hardly changed (Figs. \ref{fig:SPM}f and \ref{fig:SPM}g).
The results in Fig. \ref{fig:SPM} suggest that microparticle motion was confined along line PQ (Figs. \ref{fig:SPM}b and \ref{fig:SPM}e) when the microparticles were sparsely distributed; i.e., net transport along the $y$-axis was not observed. In addition, the results suggest that the line PQ was the stable position for the microparticles.

\begin{figure*}[t]
\centering
\includegraphics[scale=0.25]{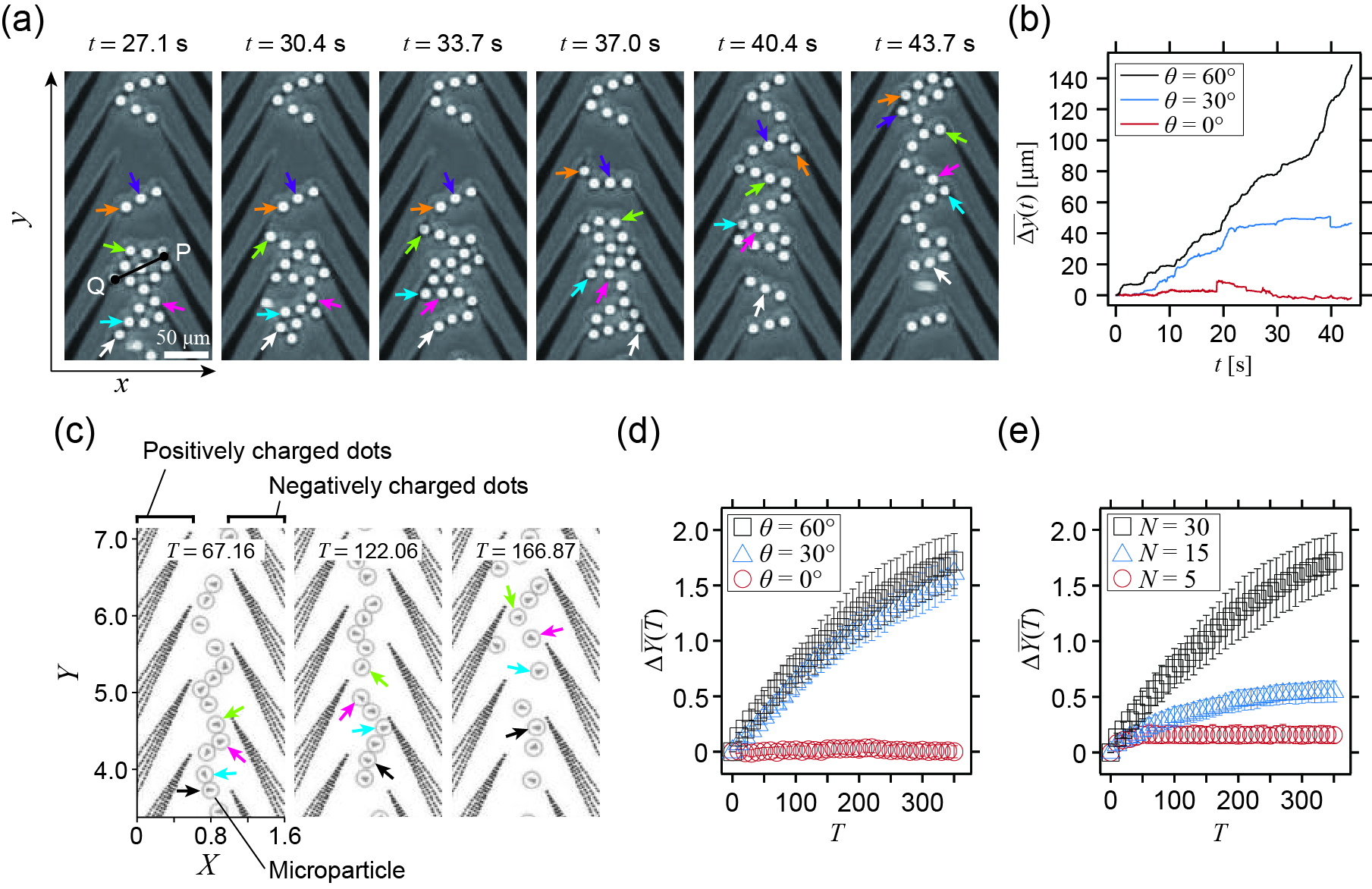}
\caption{
Collective transport of microparticles generated by crowding. (a) Snapshots of the collective transport observed in the experimental system. The microparticles were transported toward the $y$-direction. The arrow colors identify each microparticle. An electrode with $\theta = 60^{\circ}$ was used. Line PQ is indicated in the snapshot at $t = 27.1$ s. (b) Time variations of $\overline{\Delta y}(t)$ (black line: $\theta = 60^{\circ}$, blue line: $\theta = 30^{\circ}$, and red line: $\theta = 0^{\circ}$). Increase in $\overline{\Delta y}(t)$ was clearly observed in cases where $\theta = 60^{\circ}$ and $\theta = 30^{\circ}$.
(c) Snapshots of the collective transport observed in the numerical simulation. The arrows colors identify each microparticle. The charged dots are patterned along the electrode with $\theta = 60^{\circ}$.
The details are shown in Fig. S6.
(d) Dependence of $\overline{\Delta Y}(T)$ on the asymmetry of the potential (black square: $\theta = 60^{\circ}$, blue triangle: $\theta = 30^{\circ}$, and red circle: $\theta = 0^{\circ}$). $N = 30$. (e) Dependence of $\overline{\Delta Y}(T)$ on the number of particles (black square: $N = 30$, blue triangle: $N = 15$, and red circle: $N = 5$). $\theta = 60^{\circ}$. In (d, e), the error-bars indicate the standard deviation.
}
\label{fig:CM}
\end{figure*}

\vspace {0.1in}
\noindent \textbf{Collective microparticle transport generated during particle crowding.} Fig. \ref{fig:CM}a displays the snapshots of the collective transport of microparticles between the electrodes at $\theta = 60^{\circ}$ for each $t$ (Movie S3).
The arrow colors identify each microparticle. Here, we define the mean displacement for all the microparticles as $\overline{\Delta y}(t)= \langle y(t)-y(0) \rangle $ (black line in Fig. \ref{fig:CM}b).
Unlike the case of the sparse microparticles (Fig. \ref{fig:SPM}), some microparticles were transported along the $y$-axis.
The collective transport was also observed when $\theta = 30^{\circ}$ (blue line in Fig. \ref{fig:CM}b, Fig. S5a and Movie S4).
In contrast, a distinct increase in $\overline{\Delta y}(t)$ was not observed when the electrode with $\theta = 0^{\circ}$ was used (red line in Fig. \ref{fig:CM}b, Fig. S5b and Movie S5).
We speculate that the collective transport under an asymmetric electric field was generated by the electrostatic interaction and collision among the crowding microparticles.
When the microparticles were sparsely dispersed between the electrodes (Fig. \ref{fig:SPM}), the microparticle motion was confined between points P and Q, which was a stable position; hence, transport along the $y$-axis was not observed.
However, during crowding (Fig. \ref{fig:CM}a), all the microparticles could not exist on line PQ (snapshot at $t = 27.1$ s in Fig. \ref{fig:CM}a) due to the microparticle volume. In the periphery of line PQ, these microparticles interacted and received randomly directed forces.
Eventually, when the electric field was asymmetric, some microparticles were stochastically transported along the $y$-axis due to the locally biased potential (black and blue lines in Fig. \ref{fig:CM}b, and Movies S3 and S4).
On the other hand, when the electric potential was symmetric, although microparticle interaction was observed, net transport was not observed (red line in Fig. \ref{fig:CM}b, Movie S5).

To support this hypothesis, we performed numerical simulations considering the forces exerted on the microparticles.
A microparticle receives an electrostatic force, a dielectric force, and a force through the interaction with other microparticles.
Then, the equation of motion of the $\it{i}$-th microparticle is described as follows:
\begin{equation}
  \eta \dfrac{\mathrm{d}\bm{R}_i}{\mathrm{d}T} = \sum_{j} \sum_{k}(\bm{F}_{\mathrm{EP}{ijk}}+\bm{F}_{\mathrm{DEP}ijk}) + \sum_{i^{\prime}}(\bm{F}_{\mathrm{PP}ii^{\prime}}), \label{eq:2DME}
\end{equation}
where $\eta$ is the viscous constant, $\bm{R}_i = (X_i, Y_i)$ is the position of the $\it{i}$-th microparticle in a dimensionless $X$-$Y$ coordinate system, and $T$ is the time.
Here, we represent the electrodes as a patterned cluster of fixed dots with charge; i.e., the positively charged dots and the negatively charged dots are patterned along the shape of the electrodes (Fig. S6a).
$\bm{F}_{\mathrm{EP}{ijk}}$ and $\bm{F}_{\mathrm{DEP}ijk}$ represent the electrostatic and dielectrostatic forces between the $\it{i}$-th microparticle and the $j$-th positively charged dot and $k$-th negatively charged dot representing electrodes (see eq.\ref{eq:EP} and eq.\ref{eq:DEP} in the Materials and Method section), and $\bm{F}_{\mathrm{PP}ii^{\prime}}$ describes the interaction force between the $i$-th and $i^{\prime}$-th microparticles attributed to the electrostatic force and collision (see eq.\ref{eq:INT}  in Materials and Method section).
In addition, each microparticle has charge, and the charge changes on contact with the charged dots or other microparticles.
The details of this model are explained in the Materials and Methods section.
Fig. \ref{fig:CM}c displays the images of the numerical simulation using eq.\ref{eq:2DME} under the condition that the positive and negative dots are patterned along the electrode with $\theta = 60^{\circ}$ (Movie S6).
Each image is a cropped snapshot from an original simulation window (Fig. S6b).
Each arrow in Fig. \ref{fig:CM}c indicates that some microparticles are transported with time $T$.
Fig. \ref{fig:CM}d depicts the plot of the mean displacement of all the microparticles, $\overline{\Delta Y}(T) = \langle Y(T) - Y(0) \rangle$.
The black squares and blue triangles denote $\overline{\Delta Y}(T)$ under $\theta = 60^{\circ}$ and $30^{\circ}$, respectively, indicating that net transport is generated when the potential is asymmetric (Fig. \ref{fig:CM}c, Fig. S7a and Movies S6 and S7).
On the other hand, when the potential is symmetric ($\theta = 0^{\circ}$) (Fig. S7b and Movie S8), there is no significant change in $\overline{\Delta Y}(T)$ as indicated by the red circles in Fig. \ref{fig:CM}d, which is consistent with the experimental results (red line in Fig. \ref{fig:CM}b).
Fig. \ref{fig:CM}e represents the dependence of the net transport on the number of particles $N$ in the simulation. While an increase in $\overline{\Delta Y}(T)$ is clearly observed when $N = 30$ (black squares in Fig. \ref{fig:CM}e), when $N = 15$ and 5 (blue triangles and red circles in Fig. \ref{fig:CM}e, respectively), $\overline{\Delta Y}(T)$ is less than half of $\overline{\Delta Y}(T)$ for $N = 30$, indicating that the crowding of particles induces transport.
These results qualitatively agree with our experimental results. The cases when $N = 5$ and 30 might correspond to the experimental observation of sparse microparticles (Fig. \ref{fig:SPM}) and crowding (Fig. \ref{fig:CM}a), respectively, and $N = 15$ may correspond to an intermediate situation.
Thus, we conclude that the collective transport in our system (Fig. \ref{fig:CM}a) can be attributed to the asymmetricity of the potential and the crowding of microparticles.

\begin{figure}[!t]
\centering
\includegraphics[scale=0.2]{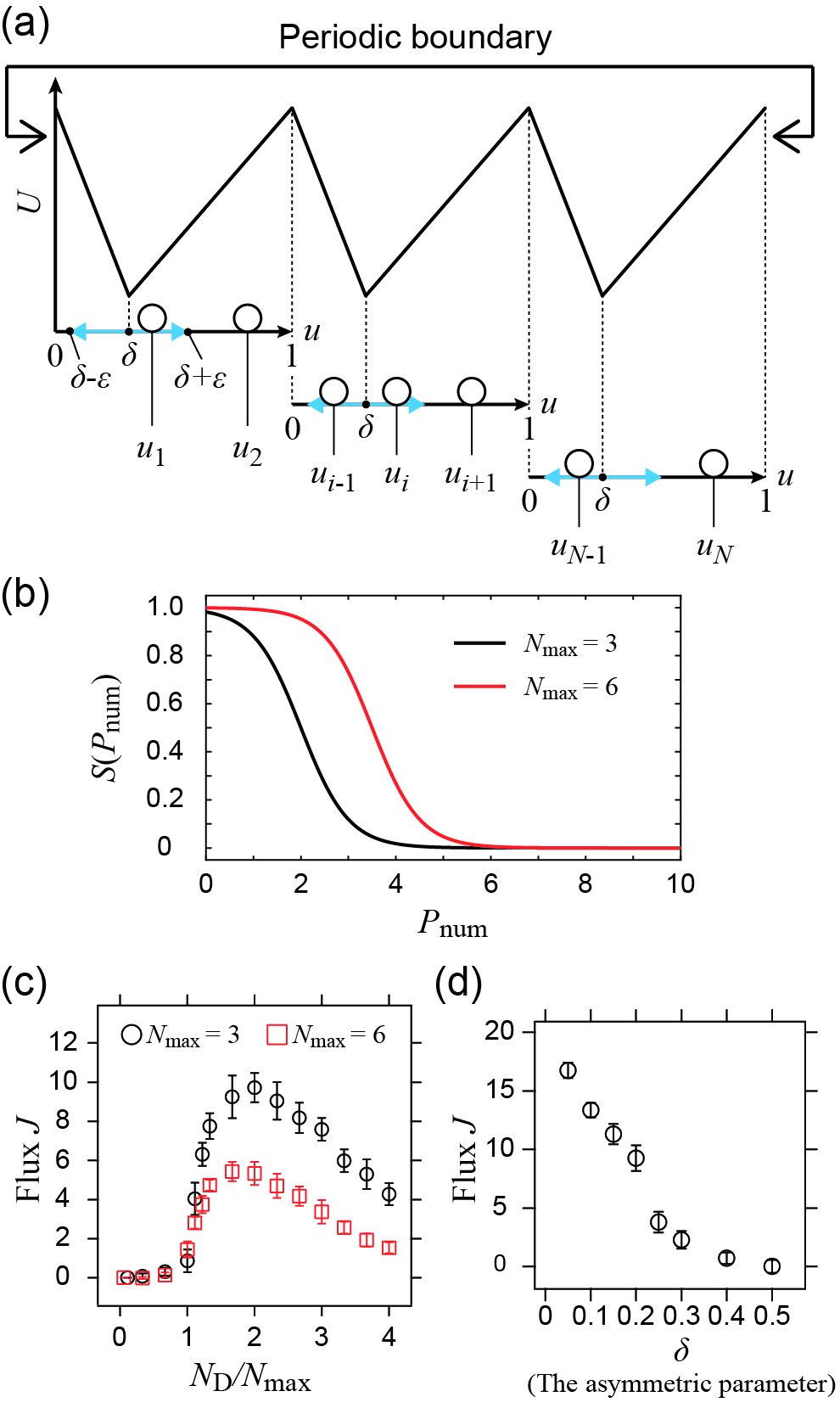}
\caption{
Modelling and calculation of the collective transport generated by crowding.
(a) Schematic of the one-dimensional potential.
Our system comprises a one-dimensional line that includes three periodic potentials connected by the periodic boundary condition.
The potential has a period of unity and its asymmetricity is determined by the asymmetricity parameter $\delta$.
In our model, the force due to the potential is weakened depending on the number of microparticles existing in the periphery of the minimum, indicated by blue arrows ($\delta-\epsilon < u < \delta+\epsilon$).
In all the simulations, $\epsilon = 0.2$.
(b) Sigmoid function as a function of $P_\mathrm{num}$.
When there are sufficient microparticles in the minimum, potential $U$ is weakened by the multiplication of this function.
Black line: $N_\mathrm{max} = 3$ and $\gamma = 2$. Red line: $N_\mathrm{max} = 6$ and $\gamma = 1$.
(c) Dependence of $J$ on $N_\mathrm{D}/N_\mathrm{max}$.
$J$ increases when $N_\mathrm{D}/N_\mathrm{max}$ exceeds unity, and has a peak at $N_\mathrm{D}/N_\mathrm{max} \sim 2$.
After the peak, $J$ gradually decreases. $\delta = 0.2$. Black circle: $N_\mathrm{max} = 3$. Red square: $N_\mathrm{max} = 6$.
(d) Dependence of $J$ on $\delta$.
$J$ decreases with the increase in $\delta$. $N = 30$ and $N_\mathrm{max} = 3$.
}
\label{fig:AM}
\end{figure}

\vspace {0.1in}
\noindent \textbf{Modeling of the collective transport generated by crowding.} Our experimental and simulation results indicate that microparticle crowding generates net mass transport under asymmetric periodic potential. Because the possible number of microparticles confined at minimum potential is limited, some of the microparticles are stochastically pushed out along the $y$-axis, resulting in microparticle transport.
This manner of transport is the same as that in the previous theoretical study that depicts the mass transport of microparticles under an asymmetric potential, where spatial overlapping is forbidden \cite{derenyi1995cooperative}.
From another point of view, it can also be argued that the crowding of microparticles forces the field "flashing".
In this study, we modeled the collective transport as the system whose potential “flashes” depending on the number of particles in the period.
The calculations were performed along a one-dimensional line that included three periodic potentials with a period length of unity under a periodic boundary condition (Fig. \ref{fig:AM}a).
The nondimensional model is described as
\begin{equation}
  \zeta\dfrac{\mathrm{d}u_i}{\mathrm{d}\tau} = -\dfrac{\mathrm{d}U(u_i)}{\mathrm{d}u}\cdot S(P_{\mathrm{num}}) + \xi(\tau),
\end{equation}
where $\zeta$ represents the viscous constant, $u_i$ is the position of the $i$-th microparticle, $\tau$ is the time, and $\xi$ is a random force that follows a Gaussian distribution.
A periodic boundary condition is assumed; i.e., $0 < u < 1$.
$U$ is the periodic potential as shown in Fig. \ref{fig:AM}a, with the following form:
\begin{align}
  U(u_i)&= \left\{ \begin{array}{l}
  \dfrac{U_\mathrm{max}}{\delta}(\delta-u_i) \quad (0 \leq u_i \leq \delta) \\ [5pt]
  \dfrac{U_\mathrm{max}}{1-\delta}(u_i-\delta) \quad (\delta < u_i < 1)
  \end{array}
  \right.,
\end{align}
where $\delta$ is the asymmetricity parameter indicating the position where the potential is minimum and $U_\mathrm{max} > 0$ is the maximum value of the potential.
In our model, $U(u_i)$ is weakened depending on a sigmoid function (Fig. \ref{fig:AM}b) described as
\begin{equation}
  S(P_\mathrm{num}) = \dfrac{1}{1+\exp\left(\gamma\left(P_\mathrm{num}-\dfrac{N_{\mathrm{max}}+1}{2}\right)\right)},
\end{equation}
where $P_\mathrm{num}$ is the number of microparticles existing in the periphery of the minimum potential, represented as $\delta-\epsilon < u < \delta+\epsilon$ (Fig. \ref{fig:AM}a).
The range of the periphery of the minimum potential is determined by $\epsilon$; in all the simulations, $\epsilon = 0.2$.
$\gamma$ is the gain, and $N_{\mathrm{max}}$ is the maximum number of microparticles that can occupy the minimum potential, $\delta-\epsilon < u < \delta+\epsilon$.
To evaluate the transport, we define the flux $J = (1/N) \sum_{\tau=0}^{S}j(\tau)$, where $N$ is the total number of microparticles in the simulation, and $S$ is the final time.
$j(\tau)$ is defined as $j(\tau) = \sum_{i=0}^{N} j_i(\tau)$, where $j_i(\tau)$ is a variable that characterizes the transport of the $i$-th microparticle at time $\tau$.
When the $i$-th microparticle moves into the subsequent period of the potential on the left , $j_i(\tau) = 1$; when it moves into the right, $j_i(\tau) = -1$; when it remains at the same period, $j_i(\tau) = 0$.

Fig. \ref{fig:AM}c displays a plot of $J$ against $N_{\mathrm D}/N_{\mathrm{max}}$.
$N_{\mathrm D}$ denotes the number of microparticles per the period of the asymmetric potential ($N_{\mathrm D} = N/3$).
$J$ is small when $N_{\mathrm D}/N_{\mathrm{max}}$ is small $(N_{\mathrm D}/N_{\mathrm{max}} < 1)$, and it begins to increase significantly when $N_{\mathrm D}/N_{\mathrm{max}} = 1$.
This nonlinear response is observed in both cases, where $N_{\mathrm{max}} = 3$ and $6$, respectively (Black circles and red squares in Fig. \ref{fig:AM}c, respectively).
$N_{\mathrm D}/N_{\mathrm{max}} = 1$ indicates that a sufficient number of microparticles exist to weaken the potential force at all the periodic potentials.
These results well reproduce our experimental results.
The small $J$ at small $N_{\mathrm D}/N_{\mathrm{max}}$ corresponds to the back-and-forth motion and the alignment on line PQ in the sparsely populated areas, and the increase in $J$ in $1 < N_{\mathrm D}/N_{\mathrm{max}} < 2 $ corresponds to the collective motion generated by crowding.
When $N_{\mathrm D}/N_{\mathrm{max}} > 2$, $J$ gradually decreases.
Additionally, it should be noted that $J$ of $N_{\mathrm{max}} = 6$ is always lower than that of $N_{\mathrm{max}} = 3$.
We speculate that the decrease in $J$ may be caused by excessive crowding.
When there are too many particles in the system, the potential force is weakened at all the periods of the potential, and the motions of the particles are randomized, causing a decrease in $J$.
This nonlinear relationship between the particle number and transport efficiency can be beneficial, for example, to estimate the appropriate sample density in an actual transport system.

Fig. \ref{fig:AM}d depicts the plot of $J$ against $\delta$, indicating that asymmetricity in the periodic potential is essential for transport.
In the presence of asymmetricity ($0.05 \leq \delta < 0.5$), transport is always generated; however $J$ decreases with the decrease in the amplitude of the asymmetricity.
Further, when $\delta = 0.5$, the potential is symmetric and transport is not generated.
Similar phenomena are experimentally observed, as shown in Fig. \ref{fig:CM};
transport is not observed when the electric field is symmetric ($\theta = 0^{\circ}$), whereas collective transport is observed when $\theta = 60^{\circ}$.

\vspace {0.5in}

\leftline {\large \textbf{Conclusion}}
In this study, we demonstrated the collective transport generated by the crowding of microparticles under a periodic asymmetric potential.
When the microparticles were sparsely distributed, they exhibited back-and-forth motion (Figs. \ref{fig:SPM}b-\ref{fig:SPM}d) and stable alignment on line PQ (Figs. \ref{fig:SPM}e-\ref{fig:SPM}g); the microparticles were confined in the minimum of the periodic potential.
On the other hand, when the microparticles were crowded and the field was asymmetric, they moved into the next period of the potential due to electrostatic and collision interaction (Figs. \ref{fig:CM}a and \ref{fig:CM}b).
The results of the two-dimensional simulation agree well with the experimental results indicating that both crowding and an asymmetric potential are necessary for mass transport (Figs. \ref{fig:CM}c-\ref{fig:CM}e).
Finally, we constructed an abstract model of the collective transport generated by crowding.
In the model, the amplitude of the potential is assumed to vary depending on the number of microparticles.
This abstract and simpler model succeeds in depicting the generation of mass transport observed in our experimental system (Figs. \ref{fig:AM}c and \ref{fig:AM}d).
This work verifies the theoretically expected phenomenon of the generation of ratchet transport by microparticle interaction.
The generated collective ratchet transport is a transport technique harnessing self-organized collective motion through individual interaction \cite{reichhardt2017ratchet,vicsek2012collective,sumino2012large,bricard2013emergence}.
We believe that our study can lead to the development of transport techniques on a micrometer scale with sophisticated functions such as changes in the transport direction, and sorting in accordance with the particle shape and size
\cite{hayakawa2016complex,habasaki2015vertical,hernandez2007colloidal}.


\vspace {0.5in}

\leftline {\large \textbf{Materials and Methods}}
\noindent \textbf{Preparation and experimental setup:} Polystyrene microparticles (without coating, size: 20 $\mu$m, 01-00-204S, micromod, Germany) dispersed in liquid paraffine (K-350, Kaneda, Japan) with 0.1$\% \mathrm{(w/w)}$ Span80 (Sorbitan Monooleate, Tokyo Chemical Industry, Japan) were sonicated for 60 min, and filled in a PDMS chamber (height: approximately 1 mm) which was placed on the glass slide with the electrode. The open top was sealed with cover glass (1.8 cm $\times$ 1.8 cm, Matsunami Glass, Japan) to prevent convectional flow in the liquid paraffin. Static DC voltage was then applied to the electrode using a constant-voltage power supply (GPR-30H100, GW Instek, Japan). The motions of the microparticles were observed using an inverted microscope (CKX-41, Olympus, Japan) and recorded using a digital camera (EOS D60, Canon, Japan) at a frame rate of 30 fps. The recorded video was analyzed using image processing software, Image J (National Institute of Health, U.S.).

\vspace{0.1in}
\noindent \textbf{Two-dimensional numerical simulation:} Each force in eq.\ref{eq:2DME} is described as follows.
\begin{equation}
  \bm{F}_{\mathrm{EP}{ijk}} = aQ_i(s_i) \left( \dfrac{\bm{d}_{+ij}}{|\bm{d}_{+ij}|^3}-\dfrac{\bm{d}_{-ik}}{|\bm{d}_{-ik}|^3} \right). \label{eq:EP}
\end{equation}
\begin{equation}
  \bm{F}_{\mathrm{DEP}{ijk}} = b\left( \dfrac{\bm{d}_{+ij}}{|\bm{d}_{+ij}|^6}+\dfrac{\bm{d}_{-ik}}{|\bm{d}_{-ik}|^6} \right). \label{eq:DEP}
\end{equation}
\begin{equation}
  \bm{F}_{\mathrm{PP}ii^{\prime}} = -c\dfrac{\bm{d}_{ii^{\prime}}}{|\bm{d}_{ii^{\prime}}|^3}m_i m_{i^{\prime}} - d(2R_{c}-|\bm{d}_{ii^{\prime}}|)\dfrac{\bm{d}_{ii^{\prime}}}{|\bm{d}_{ii^{\prime}}|}. \label{eq:INT}
\end{equation}
\begin{equation}
  a, b, c, d = \mathrm{const.} \label{eq:CON}
\end{equation}
$\bm{F}_{\mathrm{EP}{ijk}}$ in eq.\ref{eq:EP} is described as the force inversely proportional to the square of the distance from the electrode, which corresponds to the electrostatic force. $\bm{d}_{+ij}$ and $\bm{d}_{-ik}$ are the distances from the $\it{i}$-th microparticle to the positive and negative dots, respectively. $Q_i(s_i) = m_i\beta^{s_i}$ is defined as the charge of the $\it{i}$-th microparticle. $Q_i(s_i)$ exponentially dissipates with elapsed time $s_i$ on contact with either dots, and $\it{m_i}$ is a discreate variable representing the sign of the charge ($m_i = -1, 1$); $\beta < 1$ is the dissipation parameter.
The contact of the $\it{i}$-th microparticle of radius $\it{R_c}$ with the positive and negative dots is determined by $|\bm{d}_{+ij}| \leq R_c$ and $|\bm{d}_{-ik}| \leq R_c$.
After contact, $\it{m_i}$ is updated to $m_i = -1$ (the case where contact occurs with the positive dot) and $m_i = 1$ (the case where contact occurs with the negative dot), and $s_i$ is reset to $s_i = 0$.
$\bm{F}_{\mathrm{DEP}{ijk}}$ in eq.\ref{eq:DEP} is described as the force inversely proportional to the fifth power of $\bm{d}_{+ij}$ and $\bm{d}_{-ik}$, which corresponds to the dielectrophoretic force.
Here, we assume that the dielectric constant of the polystyrene microparticle is larger than that of liquid paraffin; therefore, $\bm{F}_{\mathrm{DEP}{ijk}}$ acts in the direction in which the microparticles are attracted to the electrode.
$\bm{F}_{\mathrm{PP}ii^{\prime}}$ in eq.\ref{eq:INT} describes the interaction among the particles, attributed to the electrostatic force and collision.
The first term on the right of eq.\ref{eq:INT} represents the electrostatic force acting between the $i$-th and $i^{\prime}$-th microparticles.
$\bm{d}_{ii^{\prime}}$ represents the distance between the $i$-th and $i^{\prime}$-th microparticles.
The second term is the repulsive force that prohibits the overlapping of the $i$-th and $i^{\prime}$-th microparticles, and it repels them when $|\bm{d}_{ii^{\prime}}| \leq 2R_c$.
Furthermore, we consider the charge transfer among particles parallel to the interaction force (eq.\ref{eq:INT}).
When the microparticles contacts, the charge of contacting microparticles change to $\langle Q_i(s_i) \rangle_{|\bm{d}_{ii^{\prime}}| \leq 2R_c}$, meaning the average of the contacting microparticle charge. Additionally, $s_i$ is reset to $s_i = 0$.
The equation of motion and the force formulas were normalized into a dimensionless system.
In all the simulations, $\eta$, $R_c$, $\beta$, and $(a, b, c, d)$ were set to 1.0, 0.091, 0.996, and (0.01, 0.000015, 0.01, 50.0), respectively, and the distance between the electrodes ($d_\mathrm{h}$ in the experiments) was 0.456. The calculation was performed for 50000 steps.

\vspace{0.1in}
\noindent \textbf{Calculation of the abstract one-dimensional model:} The equation of motion was normalized into a dimensionless system.
In the simulations depicted in Fig. \ref{fig:AM}c, $\gamma$ = 2 when $N_{\mathrm{max}} = 3$ (Black circles), and $\gamma$ = 1 when $N_{\mathrm{max}} = 6$ (red squares). In the simulations depicted in Fig. \ref{fig:AM}d, $\gamma$ = 2 and $N_{\mathrm{max}} = 3$.
In these simulations, $\epsilon = 0.2$, and the calculations were performed for 10,000,000 steps.

\vspace{0.1in}
The methods for the fabrication of sawtooth-shaped electrode and the calculation of electric potential are described in Supplementary Note.

\vspace {0.5in}

\leftline {\large \textbf{Acknowledgment}}
We thank Prof. K. Yoshikawa (Doshisha Univ.) and Dr. M. Morita (AIST) for the fruitful discussions. This work was supported by JSPS KAKENHI Grant Numbers JP17H01813 and JP18K19834.

\clearpage
\bibliography{ref.bib}

\end{document}